**Title**: Phase-type distributions for studying variability in resistive memories

**Authors:** C. Acal[1], J.E. Ruiz-Castro[1], A. M. Aguilera[1], F. Jiménez-Molinos[2], J.B. Roldán[2]

**Address**

[1]Departamento de Estadística e Investigación Operativa and IEMath-GR. Universidad de Granada. Facultad de Ciencias. Campus de Fuentenueva s/n, 18071 GRANADA, Spain.

[2]Departamento de Electrónica y Tecnología de Computadores. Universidad de Granada. Facultad de Ciencias. Avd. Fuentenueva s/n, 18071 GRANADA, Spain.

Correspondence author e-mail: aaguiler@ugr.es

**Abstract**

A new statistical approach has been developed to analyze Resistive Random Access Memory (RRAM) variability. The stochastic nature of the physical processes behind the operation of resistive memories makes variability one of the key issues to solve from the industrial viewpoint of these new devices. The statistical features of variability have been usually studied making use of Weibull distribution. However, this probability distribution does not work correctly for some resistive memories, in particular for those based on the $Ni/HfO_2/Si$ structure that has been employed in this work. A completely new approach based on phase-type modelling is proposed in this paper to characterize the randomness of resistive memories operation. An in-depth comparison with experimental results shows that the fitted phase-type distribution works better than the Weibull distribution and also helps to understand the physics of the resistive memories.

*Key words*

# 1. INTRODUCTION

In the context of applications for non-volatile memories, several emerging technologies are gaining momentum in the electronic industry. Among the new devices considered, both at industry and academia, RRAMs have shown an incomparable potential because they have good scalability, low power operation, fast speed and outstanding possibilities for fabrication in the current CMOS technology [1, 2, 3, 4, 5, 6].

The physics and internal properties of these devices have been studied by means of profound experimental studies [1, 7, 8, 9], and modelling and simulation studies [10, 11, 12, 13, 14, 15, 16]. A unified mathematical framework capable of describing and simulating the complex and interplaying electro-thermo-chemical processes that occur in this type of new emerging technologies in semiconductor device industry was recently developed [17]. Nevertheless, there are issues, such as variability, that have to be addressed prior to RRAM massive industrialization.

RRAMs operation is based on the stochastic nature of resistive switching (RS) processes that, in most cases, create (set process) and rupture (reset process) a conductive filament that changes drastically the device resistance [1, 5, 18]. There is a great need to analyse the statistics behind RRAM variability that is translated to different resistances, voltages and currents related to set and reset processes for each RS cycle (a cycle consist of a set process followed by a reset process) within a long series of cycles. Because of this, the choice of the right statistical model to describe the distribution of switching parameters (forming, set and reset voltages) is a critical requirement for RRAMs that ensures a robust design of the circuit and reliable data storage unit.

The usual statistical analysis performed on experimental data measured in these devices makes use of the Weibull distribution (WD) [1, 19, 20]; nevertheless, sometimes its fit with experimental data is not very accurate. In the next section, it will be shown that WD does not work correctly for the devices under consideration in this manuscript. In fact, recent dielectric breakdown studies have shown that the WD does not describe the stochastic trends well enough, more so in downscaled structures at the low and high percentile regions. The validity of a defect clustering model for RRAM switching parameters was recently examined [21].

Therefore, another statistical approach is needed. Apart from an accurate statistical description of experimental data, the interpretation of the parameters extracted by the application of the statistical analysis can shed new light to the variability issue and the physics behind RRAM operation.

In order to deepen on this issue, a thorough analysis of the statistical properties of RRAM variability is performed. To do so, phase-type distributions (PHDs) will help us to analyse the possible intermediate states of degradation in the conductive filament destruction processes that lead to a RRAM high resistivity state (the rupture of the conductive filament isolates the electrodes and therefore the RRAM resistance increases).

Phase-type distributions, which were introduced and analyzed in detail by Neuts [22, 23, 24], constitute a class of non-negative distributions that makes it possible to model complex problems with well-structured results, thanks to its matrix-algebraic form. Due to their valuable properties, many varieties of this class of distributions have been considered in diverse branches of science and engineering and applied in reliability studies. Particular cases of PHDs are the exponential, Erlang, generalized Erlang, hyper-geometric and Coxian distributions, among others. In fact, not only very well-known probability distributions are PHDs but also any nonnegative probability distribution can be approximated as needed taking into consideration that the PHD class is dense in the set of probability distributions on the nonnegative half-line [25]. The more essential and important properties of PHDs were reviewed in a recent study [26].

As reported below, the versatility and advantages of the PHD will come up to allow a better analysis of RRAM variability. In this context, it will be shown that the Erlang distribution (ED) (a particular PHD) works much better than WD to describe the experimental data under consideration in this manuscript. The physical interpretation of the fitted parameters from the PHD modelling will shed light on the explanation of RRAM variability.

The fabricated devices and measurement process are described in Section 2, the new statistical approach for modelling RRAMs variability is given in Section 3 and the main results and discussion in Section 4. Finally, the conclusions are given in Section 5.

## 2. DEVICE DESCRIPTION AND MEASUREMENT

The devices employed in this manuscript are unipolar $Ni/HfO2/Si$-based RRAMs. The fabrication details were given in [27]. A HP-4155B semiconductor parameter analyser was used in the measurement process that consisted of a long series of RS cycles under ramped voltage stress. The Si substrate (bottom electrode) was grounded and a negative voltage was applied to the Ni (top electrode), although for simplicity we have assumed the absolute value of the applied voltage henceforth [27].

Three reset current versus voltage curves of the series of resistive switching cycles (2749 cycles) measured are shown in Figure 1. In the curves plotted, a sudden current drop can be seen corresponding with the rupture of the conductive filament (highlighted as reset point [1, 5, 27, 14, 15]) that connects the electrodes (in this respect, the conductive filament works as a fuse). The corresponding voltages and currents are known as reset voltages and currents respectively [1, 18], they have been explicitly shown in Figure 1 for the sake of clarity. The reset voltage determination has been performed by detecting a 50% current variation at the reset point (when the sudden current drop takes place). Others methods have been proposed in the literature [18]; nevertheless, this one worked well for the devices under consideration here.

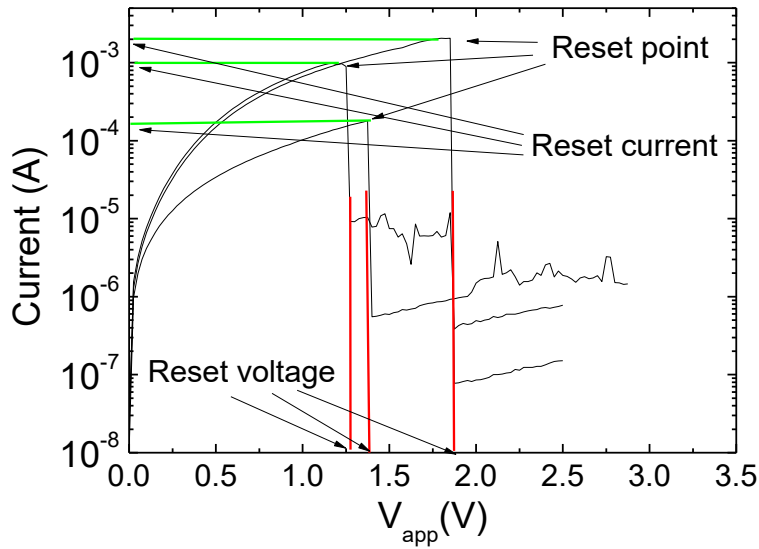

**Figure 1.** Experimental current versus applied voltage (shown in black lines) for three curves of a 2749 series of continuous resistive switching cycles, including set and reset curves. The reset point and the corresponding reset voltages ($V_{reset}$, indicated by vertical red lines) and reset currents ($I_{reset}$, indicated by horizontal green lines) are shown for clearness.

As noted in the introduction, the $V_{reset}$ and $I_{reset}$ distributions (as well as $V_{set}$ and $I_{set}$ distributions) have been subject of a deep statistical analysis for many different RRAMs [20, 28]. The WD was employed to describe the statistical properties of experimental data when the reset and set parameters were extracted. The Weibull model has been successfully employed along with a geometrical cell-based model which was connected with the percolation model for oxide breakdown for $SiO_2$-based devices [29]. In addition, WD has been widely employed in the context of reliability physics and engineering [30]. Its use in the context of the statistical analysis of RRAM makes sense since it is a weakest-link type distribution, i.e., the failure of the whole is dominated by the degradation rate for the weakest element.

The cumulative distribution function for the WD is given in Equation 1 [8, 30]

$$F\left(v\right) = 1 - \exp\left(-\left(\frac{v}{\alpha}\right)^{\beta}\right) \tag{1}$$

For the devices reported above the statistical analysis based on the WD has been performed. On the one hand, the $V_{reset}$ and $I_{reset}$ for all the reset curves of the 2749 cycles under consideration in our RS series were computed [8, 31]. On the other hand, the typical Weibits, calculated as ln[-ln(1 – F)], have been obtained. If Weibits are plotted versus ln($V_{reset}$), a linear plot should be obtained if experimental data follow a WD, where the slope corresponds to the β parameter in Equation 1 (β measures the statistical dispersion [8, 30]). The results obtained for our devices are shown in Figure 2.

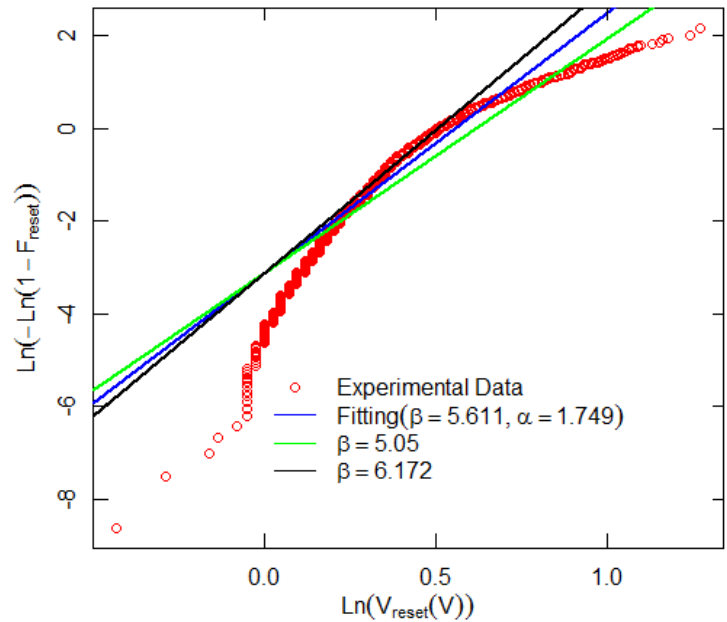

**Figure 2.** WD linear fit of $V_{reset}$ for 2749 RS cycles. The best fit is obtained with the blue line, a 10% reduction (increase) in the beta parameter was assumed in the green (black) line.

Other analytical distributions (Equation 1) were also included making use of a β parameter with a 10% variation with respect to the best fit obtained in the statistical analysis. As can be seen, the Weibits of the experimental data are not linear. Therefore, although a rough approximation could be performed in the WD context, it seems reasonable to try other distributions. We do so in the following section and we call the reader's attention to the fact that a much better fit can be obtained with a Phase-type distribution modelling. After a progressive analysis based on step-by-step estimation of phase-type distributions [22, 23], it will be shown that the ED provides the best fit. Other previous statistical analysis on RRAMs might work much better making use of the ED; however, in this manuscript the study is limited to our experimental data.

## 3. THEORY AND METHODS

As shown in previous section, the logarithm of the experimental cumulative hazard rate versus $\ln(V_{reset})$ is not linear and therefore Weibull distribution seems not to be the appropriate distribution for fitting to $V_{reset}$. Then, the aim of this section is to find out what is the best probability distribution that describes the RRAM variability.

### 3.1. Phase-type distributions

One class of non-negative probability distributions with very interesting properties that allows to describe the main associated measures in an algorithmic form and to interpret the results is the phase-type distribution class. Phase type distributions were introduced and described in detail by [22] and [23].

The flexibility of the phase-type distribution makes it a good candidate to try a better fit since it generalizes a great number of distributions. The usefulness of this distribution class has been proved in several fields such as queueing theory, renewal processes, reliability and survival [32, 33, 34, 35, 36]. In our case, we can assume that for our devices the conductive filament within the RRAM dielectric pass through different degradation stages before the rupture process takes place (absorption, in the approach we are following). At this point it seems reasonable to figure out the evolution of the conductive filament, i.e., the different stages followed in the destruction process versus $V_{reset}$.

A phase type distribution (PHD) is defined as the distribution of the lifetime up to the absorption in an absorbing Markov process (voltage up to the conductive filament failure in the RRAM context).

In the context of RRAMs an absorbing Markov process to model the voltage to the failure of the conductive filament can be assumed. The state space is given by a general number of $m$ transient degradation stages, where the probability of being initially in stage $i$ is given by $\alpha_i$ and one absorbing state, m+1, which is the conductive filament failure. In addition, the transition intensity from the transient stage $i$ to the transient stage $j$ is given by $q_{ij}$ for $i \neq j$ and if $i=j$ then $q_{ii} = -\sum_{\substack{j=1 \\ j \neq i}}^{m+1} q_{ij}$. The voltage up to failure is PHD distributed with representation $\left( \boldsymbol{\alpha}, \mathbf{T} \right)$ being $\boldsymbol{\alpha} = \left( \alpha_1, ..., \alpha_m \right)$ ; $\mathbf{T} = \left( q_{ij} \right)_{i,j=1,...,m}$ .

A PHD is a non-negative probability distribution whose cumulative distribution function is given in Equation 2

$$F\left(v\right) = 1 - \boldsymbol{\alpha} \exp\left(\mathbf{T}v\right)\mathbf{e} \ , \ v \geq 0, \tag{2}$$

where $\mathbf{e}$ is a column vector of ones with appropriate order. The density function associated to this distribution is

$$f\left(v\right) = -\boldsymbol{\alpha} \exp\left(\mathbf{T}v\right)\mathbf{T}\mathbf{e} = \boldsymbol{\alpha} \exp\left(\mathbf{T}v\right)\mathbf{T}^0 \ , \ v \geq 0,$$

with $\mathbf{T}^0 = -\mathbf{T}\mathbf{e}$ being the transition intensity vector from a transient state up to one absorbing state.

It can be seen that if $\boldsymbol{\alpha}$ is the scalar 1 and $\mathbf{T}$ is the scalar $-\lambda$, the exponential distribution is achieved. The reliability function, R($v$), describes the probability that at voltage $v$ the conductive filament is not broken, and it is given by Equation 3

$$R(v) = 1 - F(v) = \boldsymbol{\alpha}\exp(\mathbf{T}v)\mathbf{e}, \; v \geq 0. \tag{3}$$

Thus, the cumulative hazard rate is given by Equation 4

$$H(v) = -\ln(1 - F(v)) = -\ln(\boldsymbol{\alpha}\exp(\mathbf{T}v)\mathbf{e}), \tag{4}$$

and then the hazard rate is

$$h(v) = \frac{\partial H(v)}{\partial v} = \frac{f(v)}{1 - F(v)} = \frac{\boldsymbol{\alpha}\exp(\mathbf{T}v)\mathbf{T}^0}{\boldsymbol{\alpha}\exp(\mathbf{T}v)\mathbf{e}}, \; v \geq 0.$$

### 3.2. Some PHD Properties

Phase-type distributions are important not only because of their structure but also for the good properties which enable to ease the applicability and interpretation of results.

Many well known distributions, in addition to the exponential distribution mentioned above, are PHD. Next, some of these are exposed with the corresponding PHD representation.

1. The Erlang distribution $F(v) = 1 - \sum_{j=0}^{m-1} e^{-\lambda v}(\lambda v)^j / j!$ for $v \geq 0$, $m \geq 1$ and $\lambda > 0$,

$$\boldsymbol{\alpha} = (1, ..., 0, 0), \; \mathbf{T} = \begin{pmatrix} -\lambda & \lambda & & \\ & -\lambda & \ddots & \\ & & \ddots & \lambda \\ & & & -\lambda \end{pmatrix}_{m \times m}.$$

2. Hypo-exponential distribution $F(v) = 1 - \sum_{x=0}^{v} \sum_{i=1}^{m} \lambda_i e^{-\lambda_i x} \left( \prod_{\substack{j=1 \\ j \neq i}}^{m} \frac{\lambda_j}{\lambda_j - \lambda_i} \right)$ for

$v \geq 0$, $\lambda_i \neq \lambda_j$ for $i \neq j$,

$$\boldsymbol{\alpha} = (1,0,...,0),\ \mathbf{T} = \begin{pmatrix} -\lambda_1 & \lambda_1 & & \\ & -\lambda_2 & \ddots & \\ & & \ddots & \lambda_{m-1} \\ & & & -\lambda_m \end{pmatrix}.$$

3. Hyper-exponential distribution $F(v) = 1 - \sum_{i=1}^{m} \alpha_i \left(1 - e^{-\lambda_i v}\right)$ for $v \geq 0$,

$$\boldsymbol{\alpha} = (\alpha_1, \alpha_2, ..., \alpha_m),\ \mathbf{T} = \begin{pmatrix} -\lambda_1 & & & \\ & -\lambda_2 & & \\ & & \ddots & \\ & & & -\lambda_m \end{pmatrix}$$

4. Coxian distribution

$$\boldsymbol{\alpha} = (1,0,...,0),\ \mathbf{T} = \begin{pmatrix} -\lambda_1 & g_1\lambda_1 & & \\ & -\lambda_2 & g_2\lambda_2 & \\ & & \ddots & g_{m-1}\lambda_{m-1} \\ & & & -\lambda_m \end{pmatrix}$$

5. Generalized Coxian distribution

$$\boldsymbol{\alpha} = (\alpha_1, \alpha_2, ..., \alpha_m),\ \mathbf{T} = \begin{pmatrix} -\lambda_1 & g_1\lambda_1 & & \\ & -\lambda_2 & g_2\lambda_2 & \\ & & \ddots & g_{m-1}\lambda_{m-1} \\ & & & -\lambda_m \end{pmatrix}$$

In general, the following result [6] describes when a non-negative probability distribution is PHD. Then, a non-negative probability distribution is a phase-type distribution if and only if it is either the point mass at zero or; it has a strictly positive continuous density on the positive real numbers and it has a rational Laplace-Stieltjes transform with a unique pole of maximal real part.

One essential property that verifies the phase-type distribution class is that this class is dense in the set of probability distributions on the non-negative half-line [25]. In this way, PHDs can be considered general distributions with a well-structured matrix algorithmic form. On the other hand, any non-negative probability distribution can be approximated as much as desired by a phase-type-distribution.

Other properties for phase-type distributions are the closure properties. The PHD class is closed under a number of operations such as minimum, maximum and addition.

### 3.3. Estimating phase-type distributions: the EM algorithm

Fitting PHDs is a difficult optimization problem given that the representation of a PHD is highly redundant in general. One usual technique used to estimate the parameters of a PHD is the Expectation Maximization (EM) algorithm. The first EM algorithm was developed by [38] and assumed by [39].

Let $v_1,\ldots,v_n$ be a sequence of $n$ observed variable values. In our case, the value $v_i$ is the voltage up to the absorption (rupture of the filament). The set $\{v_1,\ldots,v_n\}$ includes the outcomes of $n$ independent replications of a PHD with representation $(\boldsymbol{\alpha}, \mathbf{T})$ associated to an absorbing Markov process. The likelihood function to be optimized in the EM algorithm can be expressed as

$$L(\boldsymbol{\alpha}, \mathbf{T}) = \prod_{i=1}^{m} \boldsymbol{\alpha}_i^{N_i} \prod_{i=1}^{m} e^{x_i \mathbf{T}_{ii}} \prod_{i=1}^{m} \prod_{\substack{j=1 \\ j \neq i}}^{m+1} \mathbf{T}_{ij}^{N_{ij}} \,,$$

where $N_i$ is the number of times that the Markov process started in phase $i$, $x_i$ is the total time spent in phase $i$ and $N_{ij}$ is the total observed number of jumps between both states, $i$ and $j$.

If the current estimate of the PHD is $\left( \boldsymbol{\alpha}, \mathbf{T} \right)$, then the conditional expectations of $N_i$, $x_i$ and $N_{ij}$ (E-step) are given by

$$E_{(\boldsymbol{\alpha},\mathbf{T})} \left[ N_i \right] = \frac{1}{n} \sum_{k=1}^{n} \frac{\boldsymbol{\alpha}_i \left( e^{\mathbf{T} v_k} \mathbf{T}^0 \right)_i}{\boldsymbol{\alpha} e^{\mathbf{T} v_k} \mathbf{T}^0}, \; E_{(\boldsymbol{\alpha},\mathbf{T})} \left[ x_i \right] = \frac{1}{n} \sum_{k=1}^{n} \frac{\left[ \int_0^{v_k} \left( \boldsymbol{\alpha} e^{\mathbf{T}(v_k - u)} \right)^{'} \left( e^{\mathbf{T} u} \mathbf{T}^0 \right)^{'} du \right]_{ii}}{\boldsymbol{\alpha} e^{\mathbf{T} v_k} \mathbf{T}^0},$$

$$E_{(\boldsymbol{\alpha},\mathbf{T})} \left[ N_{ij} \right] = \frac{1}{n} \sum_{k=1}^{n} \frac{\left[ \int_0^{v_k} \left( \boldsymbol{\alpha} e^{\mathbf{T}(v_k - u)} \right)^{'} \left( e^{\mathbf{T} u} \mathbf{T}^0 \right)^{'} du \right]_{ij} \mathbf{T}_{ij}}{\boldsymbol{\alpha} e^{\mathbf{T} v_k} \mathbf{T}^0}, \; E_{(\boldsymbol{\alpha},\mathbf{T})} \left[ N_{i,m+1} \right] = \frac{1}{n} \sum_{k=1}^{n} \frac{\left( \boldsymbol{\alpha} e^{\mathbf{T} v_k} \right)_i \mathbf{T}_i^0}{\boldsymbol{\alpha} e^{\mathbf{T} v_k} \mathbf{T}^0}$$

Then, the M-step results in the estimation of new parameters

$$\hat{\boldsymbol{\alpha}}_i = E_{(\boldsymbol{\alpha},\mathbf{T})} \left[ N_i \right]; \hat{\mathbf{T}}_{ij} = \frac{E_{(\boldsymbol{\alpha},\mathbf{T})} \left[ N_{ij} \right]}{E_{(\boldsymbol{\alpha},\mathbf{T})} \left[ x_i \right]} \quad , \quad i \neq j; \quad \hat{\mathbf{T}}_i^0 = \frac{E_{(\boldsymbol{\alpha},\mathbf{T})} \left[ N_{i,m+1} \right]}{E_{(\boldsymbol{\alpha},\mathbf{T})} \left[ x_i \right]}; \quad \hat{\mathbf{T}}_{ii} = -\left( \hat{\mathbf{T}}_i^0 + \sum_{\substack{j=1 \\ j \neq i}}^{m} \hat{\mathbf{T}}_{ij} \right)$$

## 4. RESULTS AND DISCUSSION

To analyze the behavior of $V_{reset}$, the classical methodology based on the Weibull distributions has been used as it can be seen in Section 2. The results are not as good as desirable. Phase-type distributions with their corresponding properties have been introduced in the section above. One interesting property of PHD is that this class of distributions is dense in the non-negative probability distributions set. Thus, PHD are going to be assumed to estimate $V_{reset}$ distribution.

The voltage up to the conductive filament failure has been fitted by considering multiple general PHDs with different stages by using the EM algorithm [38] described in Subsection 3.3. The computations have been made by using the program EMpht for fitting phase-type distributions to data [40], in addition to developing own code with the software R and using the R project for Statistical Computing [41].

Thirty PHD with $m$ transient stages, for $m = 1,...,30$, have been fitted to our data set by using the *EM*-algorithm. In total n=2749 V$_{reset}$ were observed. We have assumed any internal structure for matrix **T** (transition intensities), therefore for each one we have estimated $m(m-1)+m$ parameters. After this analysis, we have observed that the internal structure of the PHD representation depends only of one parameter, fixed $m$, for all cases. This structure can be expressed as follows

$$\boldsymbol{\alpha} = (1,0,...,0) \text{ and } \mathbf{T} = \begin{pmatrix} -\lambda & \lambda & 0 & 0 & ... \\ 0 & -\lambda & \lambda & 0 & ... \\ \vdots & \vdots & \ddots & \ddots & \vdots \\ 0 & ... & 0 & -\lambda & \lambda \\ 0 & ... & ... & 0 & -\lambda \end{pmatrix}. \quad (5)$$

It is known that a PHD with the structure described above is a Erlang distribution with representation $E(m, \lambda)$ as it can be seen in section 3.2. The cumulative distribution function of an Erlang distribution is given by

$$F(v) = 1 - \sum_{j=0}^{m-1} e^{-\lambda v} (\lambda v)^j / j!. \quad (6)$$

Thus, for the representation described in (5), both distributions: phase-type and Erlang, are equivalent, but now Erlang distribution is expressed in an algorithmic way through PH distributions. Therefore, functions (2) and (6) are the same.

Taking into account the previous results, we can conclude from the PHD analysis that the voltage up to the rupture process (V$_{reset}$) is Erlang distributed; however, the PHD structure is considered. The physical structure of the Erlang distribution and its parameters can be interpreted as follows: the conductive filament always begins in stage 1 (after a successful set process where its formation has been achieved) and it undergoes a sequential degradation thorough $m$ distinct and well differenced stages (the number of stages is characterized by parameter $m$) where the mean $V_{reset}$ in each stage is equal to $1/\lambda$ (inverse of the second parameter of the Erlang distribution).

The Erlang distribution parameters have been estimated. Table 1 shows the $\lambda$ estimates after applying the EM algorithm for different stages (whose number is described by parameter $m$).

| Number of stages | Iterations EM algorithm | LogL | Estimate $\lambda$ |
|---|---|---|---|
| 15 | 1700 | −1066.427 | 9.279325 |
| 14 | 1700 | −1096.136 | 8.660703 |
| 13 | 1300 | −1133.043 | 8.042082 |
| 12 | 1000 | −1178.317 | 7.423459 |
| 11 | 1000 | −1233.436 | 6.804838 |
| 10 | 800 | −1300.308 | 6.186217 |
| 9 | 600 | −1381.457 | 5.567595 |
| 8 | 600 | −1480.314 | 4.948974 |
| 7 | 600 | −1601.724 | 4.330352 |
| 6 | 500 | −1752.833 | 3.711730 |
| 5 | 200 | −1944.833 | 3.093108 |
| 4 | 200 | −2196.729 | 2.474486 |
| 3 | 200 | −2544.869 | 1.855864 |

**Table 1.** Parameter $\lambda$ estimated by using the EM algorithm depending on the number of stages.

The optimum value is reached for 15 stages with $\hat{\lambda} = 9.279325$. Therefore, the estimated mean $V_{reset}$ in each stage is equal to 0.1078. Finally, the mean estimated $V_{reset}$ from the beginning up to the conductive filament failure is 1.6165.

The experimental cumulative hazard rate estimated by the Erlang and Weibull distributions have been plotted and compared graphically (see Figure 3).

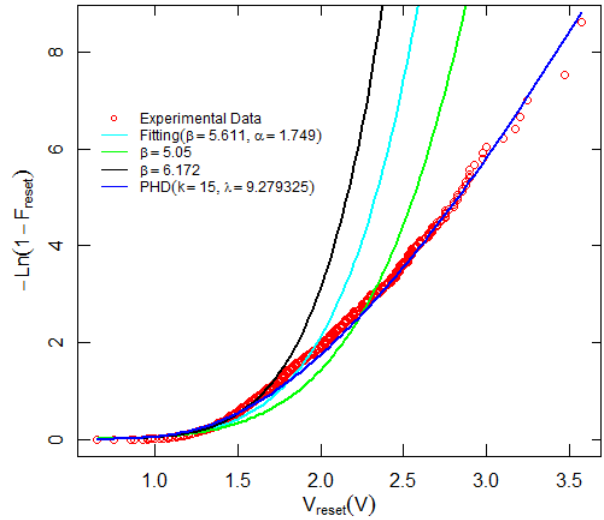

**Figure 3.** Cumulative hazard rate of $V_{reset}$ for 2749 RS cycles and the corresponding Weibull and PH distribution fit.

The best result is achieved when the Erlang distribution is considered and the accuracy of the fit is remarkable, as shown in Figure 3.

Once the statistical analysis has been developed, a detailed study of the devices considered here is developed. To do so, an analysis based on the data screened by means of the Low Resistance State (LRS) of the device resistance, R, is performed. The device LRS resistance is measured just after a set process is over, when the conductive filament is fully formed. Usually, the resistance at this point is the lowest value found all along a complete resistive switching cycle.

If $V_{reset}$ values are sorted out by considering the LRS resistance, a better fit is obtained by means of WD, although the fitting is not accurate, as can be seen in Figure 4a. In particular, for R<20 kΩ the fitting is not very good. Nevertheless, if a PHD is employed instead of a WD a reasonable accuracy is achieved. In this respect, the PHD appropriateness to deal with our experimental data is noteworthy at the sight of Figure 4b, and this is for all the resistance range under consideration.

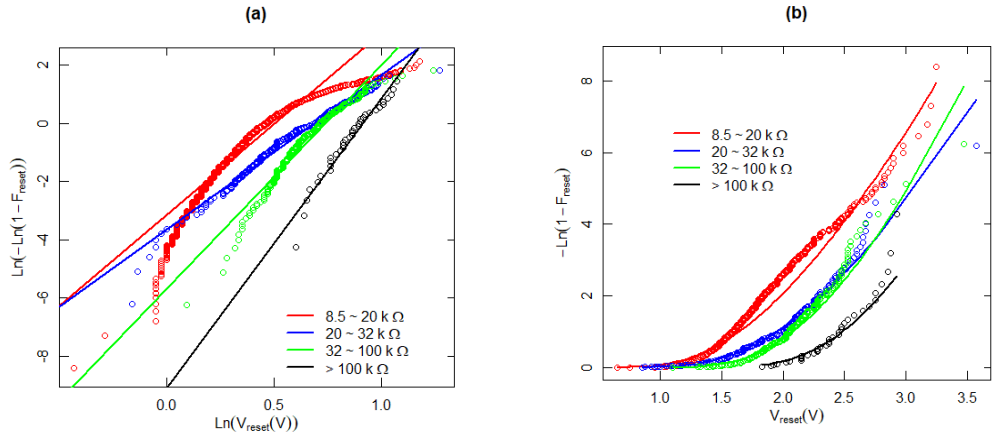

**Figure 4.** a) WD Weibits versus $Ln(V_{reset})$ for the experimental data under consideration screened for different LRS resistances are plotted in symbols. The analytical WD best fit is also shown in solid lines, b) hazard rate for the screened experimental data (symbols) and PHD (solid lines).

The reliability function, i.e., the survival function, as it is known in scientific branches not related to engineering, is interesting to analyze the statistical properties of the data we are dealing with. Since $V_{reset}$ can be considered as the failure voltage for the memories under study (RRAMs), the reliability functions portrays the probability that a memory state change will not be produced; i.e., the conductive filament will not be broken and the memory resistance state will not be switched. From another viewpoint, the reliability function describes the probability that the conductive filament will not be ruptured for voltages below the failure voltage.

The reliability function has been plotted in Figure 5 for the experimental data, WDs and PHDs. Although no distribution shows a close reproduction of the experimental values for low voltages, at medium-high voltages the PHD works better than WD and achieves a reasonably good performance.

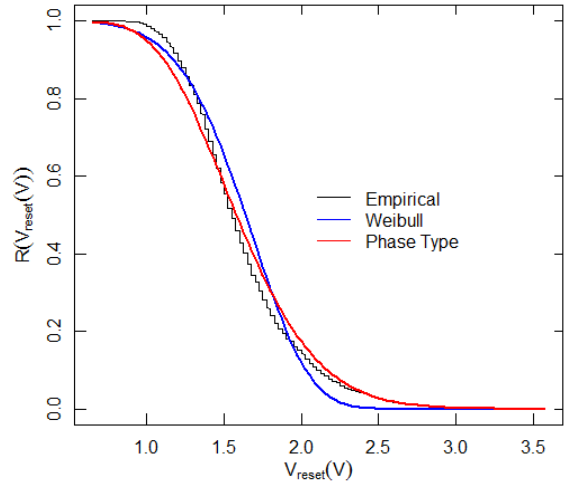

**Figure 5.** Reliability function versus $V_{reset}$ for experimental data and Weibull and Phase-type distributions.

In order to further characterize the correctness of our approach, the hazard rate function should also be considered since it describes the failure rate in a voltage interval ($V_{reset}, V_{reset}+dV_{reset}$). It could also be interpreted as the device degradation velocity at a certain voltage. This function has been plotted in Figure 6. Again, as can be seen and it was expected from previous results, PHD works better than WD.

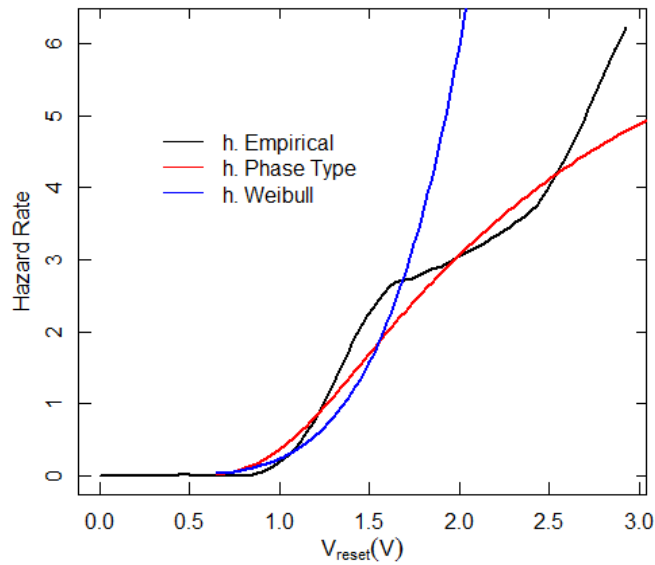

**Figure 6.** Hazard rate versus $V_{reset}$ for experimental data, phase-type and Weibull distributions.

## 5. CONCLUSIONS

The usual statistical analysis performed on RRAM experimental data in order to characterize the device variability makes use of Weibull distribution. Nevertheless, sometimes the fit obtained to measured data is not accurate. This fact suggests that other statistical distributions could work in a better manner. In this respect, a new methodology is developed in our manuscript by considering phase-type distributions to fit the $V_{reset}$ distribution, the voltage corresponding to the reset processes where RRAM conductive filaments rupture happens. The phase-type distribution class employed can be considered a general class, given that any non-negative distribution can be approximated as needed through a phase-type distribution. From the general phase-type distribution parameter estimation performed on RRAM experimental measurements, the best fit is obtained and it was found that Erlang distribution, a distribution belonging to phase-type distribution class, is particularly appropriate. The phase-type parameters were estimated from experimental data and interpreted from the physically based viewpoint. The first parameter is the number of sequential degradation stages up to the reset and the inverse of the second parameter is the mean Vreset in each stage. In addition, the fit is compared with the usual Weibull distribution to shed light on the issue.

## ACKNOWLEDGMENTS

We would like to thank F. Campabadal and M. B. González from the IMB-CNM (CSIC) in Barcelona for fabricating and providing the experimental measurements of the devices employed here. The authors thank the support of the Spanish Ministry of Economy and Competitiveness under projects TEC2014-52152-C3-2-R and MTM2017-88708-P (also supported by the FEDER program). This work has made use of the Spanish ICTS Network MICRONANOFABS.